\documentclass[journal=jctcce,manuscript=article,layout=twocolumn]{achemso}
\usepackage{amsmath}
\usepackage{graphicx}
\usepackage{amssymb}

\title{Propagators for the time-dependent Kohn-Sham equations: multistep,
  Runge-Kutta, exponential Runge-Kutta, and commutator free Magnus methods} 

\author{Adri{\'{a}}n G{\'{o}}mez Pueyo}
\email{agomez@bifi.es}
\affiliation{Institute for Biocomputation and Physics of Complex Systems, University of Zaragoza, Calle Mariano Esquillor, 50018 Zaragoza, Spain}

\author{Miguel A. L. Marques}
\affiliation{Institut f{\"{u}}r Physik, Martin-Luther-Universit{\"{a}}t Halle-Wittenberg, 06120 Halle (Saale), Germany}

\author{Angel Rubio}
\affiliation{Max Planck Institute for the Structure and Dynamics of Matter and Center for Free-Electron Laser  Science, Luruper Chaussee 149, 22761 Hamburg, Germany}
\alsoaffiliation{Center for Computational Quantum Physics (CCQ), The Flatiron Institute,  New York NY 10010}
\alsoaffiliation{Nano-Bio Spectroscopy Group, Universidad del Pa{\'{\i}}s Vasco, 20018 San Sebastián, Spain}

\author{Alberto Castro}
\affiliation{ARAID Foundation, Calle Mar{\'{\i}}a Luna, 50018 Zaragoza, Spain}
\alsoaffiliation{Institute for Biocomputation and Physics of Complex Systems, University of Zaragoza, Calle Mariano Esquillor, 50018 Zaragoza, Spain}

\date{\today}
\begin{document}

\begin{abstract}
  We examine various integration schemes for the time-dependent
  Kohn-Sham equations. Contrary to the time-dependent
  Schr{\"{o}}dinger's equation, this set of equations is non-linear,
  due to the dependence of the Hamiltonian on the electronic density.
  We discuss some of their exact properties, and in particular their
  symplectic structure. Four different families of propagators are
  considered, specifically the linear multistep, Runge-Kutta,
  exponential Runge-Kutta, and the commutator-free Magnus
  schemes. These have been chosen because they have been largely
  ignored in the past for time-dependent electronic structure
  calculations. The performance is analyzed in terms of
  cost-versus-accuracy.  The clear winner, in terms of robustness,
  simplicity, and efficiency is a simplified version of a fourth-order
  commutator-free Magnus integrator. However, in some specific cases,
  other propagators, such as some implicit versions of the multistep
  methods, may be useful.
\end{abstract}


\maketitle 

\section{Introduction}
\label{sec:introduction}

In 1984, Runge and Gross~\cite{Runge84} extended the fundamental
theorems of density-functional theory to the time-dependent case,
thereby founding time-dependent density-functional theory
(TDDFT).~\cite{Marques2012} Over the years, TDDFT has become a very
popular tool for the calculation of properties of atoms, molecules,
nanostructures, or bulk materials thanks to its favorable
accuracy/computational cost relation. It can also be used for a wide
range of applications, e.g. to calculate optical
properties,\cite{Wopperer2017} to study nuclear
dynamics,\cite{Nakatsukasa2016} charge transfer
processes,\cite{Neepa2017} electronic excitations~\cite{Rossi2017}
and ultrafast interaction of electrons with strong laser
fields,\cite{Crawford2014} to name a few.  At the core of these
simulations are the time-dependent Kohn-Sham equations (TDKS):
\begin{equation}
  \label{eq:KSE}
  \varphi_m'(t) = -i\hat{H}[n(t)](t)\varphi_m(t),\quad (m=1,\dots,N)\,,
\end{equation}
where $\hat{H}[n(t)](t)$ is the Kohn-Sham (KS) Hamiltonian, $\varphi
\equiv \{\varphi_m\}^{N}_{m=1}$ are the KS orbitals, $N$ is the number
of electrons, and $n$ is the one-electron density, obtained from
\begin{equation}
  \label{eq:KSden}
  n(\vec{r},t) = \sum_{\sigma=\uparrow,\downarrow}\sum_{m=1}^N\vert\varphi_m(\vec{r}\sigma,t)\vert^2\,.
\end{equation}
The Kohn-Sham Hamiltonian is a linear Hermitian operator, that can
have an explicit time-dependence (e.g., if the atoms are moving, or in
the presence of a laser field), and an implicit time-dependence though
the density. As the density~\eqref{eq:KSden} is written in terms of
the Kohn-Sham orbitals, Eqs.~\eqref{eq:KSE} are indeed a set of {\it
  non-linear} equations. Moreover, the KS Hamiltonian at time $t$
depends on the full history of the density at all times $t'\le t$, and
not only on its value at time $t$. These ``memory'' effects are rather
important in several circumstances (for example, for multiple
excitations), and have been extensively studied.\cite{Maitra2002}
Unfortunately, there is a lack of memory-including
exchange-correlation functionals, and the dependence on the full
history make the solution of the TDKS equations rather
complex. Therefore, almost all applications of real-time TDDFT invoke
the \emph{adiabatic approximation}, that states that the KS
Hamiltonian at time $t$ only depends on the instantaneous density at
the same time, as we already assumed in Eqs.~(\ref{eq:KSE}).

Upon discretization of the electronic Hilbert space, the TDKS
equations fall into the category of systems of initial-value
first-order ordinary differential equations (ODEs), i.e. they have the
general form:
\begin{eqnarray}
  \label{eq:general1}
  \varphi'(t) & = & f(\varphi(t),t)\,,
\\\label{eq:general2}
  \varphi(t_0) & = & \varphi_0\,.
\end{eqnarray}
Note that if history were to be considered, the TDKS equations would
no longer be an ODE system: they would belong instead to the more
general family of \emph{delay differential equations}, or
\emph{time-delay systems}.\cite{Richard2003}

Centuries of research since the early days of Newton, Euler, etc. have
produced a wide variety of numerical methods to solve
ODEs.\cite{Hairer1993,Hairer1996,Hairer2006} Any of those can in
principle be applied to the TDKS equations, but finding the most
efficient one is a difficult
task.\cite{Castro2004,Russakoff2016,Kidd2017,Dewhurst2016,Akama2015,Kolesov2016,Schaffhauser2016,Conn2015,Oliveira2015,Zhu2018}
In the following paragraphs, we make a necessarily non-exhaustive
recap of the ODE schemes that have, or have not, been tried for TDDFT
problems.

A first division can be established between \emph{one-step} and
\emph{multi-step} methods. The former provide a recipe to compute an
approximation to the solution at some time $t$ from the knowledge of
the solution at a single previous time $t-\Delta t$. The latter use
information from a number of previous steps $t-\Delta t, t-2\Delta t$,
etc. Multistep formulas have been scarcely used in the quantum
chemistry or electronic structure community, and to our knowledge
never for TDDFT calculations. Perhaps the reason is the need to store
the information about a number of previous steps, a large amount of
data for this type of problems. The most common alternatives are the
implicit and explicit formulas of Adams, and the
backward-differentiation formulas (BDFs).

For what concerns single-step methods, arguably the most used and
studied one is the family of Runge-Kutta (RK)
integrators.\cite{Butcher1987} This includes the implicit and
explicit Euler formulas, the trapezoidal (also known as
Crank-Nicolson~\cite{Crank1996}) and implicit midpoint rules, the
explicit RK4 formula (considered ``the'' RK formula since it is
perhaps the most common), the Gauss-Legendre collocation methods, the
Radau and Lobatto families, etc. Moreover, numerous possible
extensions and variations are possible: partitioned RK, embedded
formulas, use of variable time-step, extrapolation methods on top of
the RK schemes (e.g. the Gragg-Bulirsch-Stoer
algorithm~\cite{Stoer2002}), composition techniques, the linearly
implicit Rosenbrock methods, etc.  (see
Refs.~\citenum{Hairer1993,Hairer1996,Hairer2006} for a description
of these and other ideas). Once again, many of these options have
never been tested for TDDFT problems.

Linear autonomous ODE systems can also be solved directly by acting on
the initial state with the exponential of the operator that defines
the system. Quantum problems without an explicitly time-dependent Hamiltonian
belong to this class. The problem of the quantum propagator can
therefore be reduced to finding a good algorithm to compute the action
of the exponential of an operator. Various alternatives exist: a
truncation of the Taylor expansion,\cite{Flocard1978} the
Chebychev~\cite{Chen1999} and Krylov polynomial
expansions,\cite{Hochbroch1997,Frapiccini2014} Leja and Pad{\'e}
interpolations,\cite{Caliari2016} etc.

For non-autonomous linear systems (e.g. quantum problems with
time-dependent Hamiltonians), a \emph{time-ordered} exponential must
substitute the simple one. By using short-enough time-steps, however,
a constant Hamiltonian can be applied within each interval, and the
simple exponential methods mentioned above may suffice. Otherwise, one
can resort to Magnus expansions.\cite{Magnus1954} Perhaps the most
used one is also the simplest: the second-order Magnus expansion, also
known as the exponential midpoint rule. More sophisticated (higher
order) expansions require the computation of commutators of the
Hamiltonian at different times, a costly operation. Recently,
commutator-free Magnus expansions have also been
proposed.\cite{Blanes2006} Other recent options essentially based on
the exponential (and tested for TDDFT) are the non-recursive Chebychev
expansion of William Young {\em et al},\cite{Williams2015} or the
three-term recurrence of Akama {\em et al}.\cite{Akama2015}

An old-time favourite in condensed matter physics is the
split-operator formula~\cite{Feit1982}. It belongs to the wide class
of splitting techniques, whose simplest members are the
Lie-Trotter~\cite{Trotter1959} and Strang~\cite{Strang1968}
splittings. In chemistry and physics, these use the usual division of
the Hamiltonian into a kinetic and a potential part, as both can
be treated exactly in the proper representation -- the main
computational problem is then reduced to the transformation to and from
real and Fourier space. More sophisticated splitting formulas have
also been developed (see
e.g. Refs.~\citenum{Suzuki1990,Suzuki1992,Yoshida1992,Sugino1999}).

The TDDFT Hamiltonian may also be divided into a linear and a
non-linear part. The non-linear part must of course include the
Hartree, exchange, and correlation potentials. The kinetic term is
almost always included in the linear term. This is considered to be
the term responsible for the possible \emph{stiffness} of the
equations. It is difficult to give a precise definition of stiffness,
and a pragmatic one is generally accepted: ``stiff equations are
equations where certain implicit methods perform better, usually
tremendously better, than explicit ones''.\cite{Curtiss1952} Implicit
methods require the solution of nonlinear algebraic equations. Besides
outperforming explicit methods for stiff cases, some of them may also
have the advantage of preserving structural properties of the system,
such as symplecticity -- a topic that we will discuss later on. For
cases in which one part of the equation requires an implicit method,
but another part does not, the implicit-explicit (IMEX) methods were
invented.\cite{Ascher1995,Cooper1983} Another recent approach that
relies on the separation of a linear and a non-linear part are the
exponential
integrators.\cite{Hochbruck1998,Hochbruck2005,Hochbruck2006,Hochbruck2010}
There are various subfamilies: ``integrating factor'' (IF),
``exponential time-differencing'' (ETD) formulas, exponential RK,
etc. These techniques have not been tested for non-equilibrium
electron dynamics in general, or TDDFT in particular, until very
recently.\cite{Kidd2017}

An alternative that has been followed by several groups is the
transformation of the system to the adiabatic eigenbasis, or to a
closely related one (a ``spectral basis'', generally speaking). In
that appropriately chosen basis, the dimension of the system is
small, and any method can do the job. The burden of the task is then
transferred to the construction and update of the basis along the
time evolution, an operation that involves
diagonalization. Refs.~\citenum{Chen2010,Sato2014,Wang2015,Russakoff2016,Dewhurst2016}
are some recent examples, some of them based on Houston
states,\cite{Houston1940} that report notable speed-ups over
conventional methods. This result seems, however, to depend on the
type of problem, implementation details, etc.

The former list of algorithms, though long, was not exhaustive: for
example, we can also mention Fatunla's
algorithm,\cite{Frapiccini2014,Fatunla1978,Fatunla1980} or the very
recent semi-global approach of Schaefer {\em et
  al.}~\cite{Schaefer2017} based on the Chebychev propagator. It
becomes evident that the list of options is extensive, making the
identification of the most efficient, accurate, or reliable algorithm
a hard task. Some of us presented in 2003 a performance analysis of
various propagation methods for the TDKS equations;\cite{Castro2004}
it is the purpose of this Article to continue along those lines, by
investigating other promising propagation schemes and by providing
several benchmarks in order to assert their efficiency in real-world
applications. In particular, we look here at multi-step based
propagators, exponential RK integrators (along with the standard RK),
and a commutator-free version of the Magnus propagator. We implemented
these propagation schemes in our code
octopus,\cite{Marques2003,Castro2006} a general purpose
pseudopotential, real-space and real-time code.

The remaining of this article is organized in the following way: first
we study in Section~\ref{sec:theory} the theory regarding the
propagation schemes and its relation with the KS equations, paying
special attention at the issue of symplecticity. Then in
Section~\ref{sec:results} we show the benchmarks obtained for the
different propagation schemes. Finally, in
Section~\ref{sec:conclusions} we state our conclusions.

\section{Exact properties}
\label{sec:theory}

\subsection{The propagator}

If Eqs.~\eqref{eq:KSE} were linear, their solution could be written
as
\begin{equation}
	\varphi_m(t) = \hat{U}(t,t-\Delta t)\varphi(t-\Delta t)_m,\quad m=1,\dots,N\,,
\end{equation}
for some discrete time step $\Delta t$ 
(that we will consider to be constant along the evolution in this work).
The evolution operator is given by
\begin{equation}
  \label{eq:toevolution}
	\hat{U}(t,t-\Delta t)=\mathcal{T}\exp\left\lbrace-i\int^{t}_{t-\Delta t}d\tau\hat{H}(\tau)\right\rbrace\,,
\end{equation}
i.e. it is the time-ordered evolution operator. The non-linearity, however,
implies that a \emph{linear} evolution operator linking
$\varphi_m(t-\Delta t)$ to $\varphi_m(t)$ does not exist.  We may however still
assume the existence of a nonlinear evolution operator, that is
usually called a \emph{flow} in the general case,
[Eqs.~(\ref{eq:general1}) and (\ref{eq:general2})]; it is defined as
\begin{equation}
  \Phi_t(y(t-\Delta t)) = y(t)
\end{equation}
This is the object that must be approximated through some algorithm -- an algorithm that
of course takes the form of a linear operator whenever employed on
linear systems.


To choose a numerical method to propagate the TDKS equations one is
usually concerned by its performance and stability. Performance is
loosely speaking related to the computer time required to propagate
the equations for a certain amount of time. Stability, on the other
hand, is a measure on the quality of the solution after a certain
time. For linear systems (or for propagators applied to linear
systems), it is possible to give a simple mathematical definition of
stability.  A propagator is stable below $\Delta t_{\rm max}$ if, for
any $\Delta t<\Delta t_{\rm max}$ and $n>0$, $\hat{U}^n(t+\Delta t,t)$
is uniformly bounded. One way to assure that the algorithm is stable
is by making it ``contractive'', which means that $||\hat{U}(t+\Delta
t)||\le 1$. Of course, if the algorithm is unitary, it is also
contractive and hence stable; but if the algorithm is only
approximately unitary, it is better if it is contractive.  We can
also talk about unconditionally stable algorithms if their stability
is independent of $\Delta t$ and of the spectral characteristics of
$\hat{H}$.

Clearly, in many cases stability can be enhanced by decreasing
the time-step of the algorithm, i.e., by decreasing its numerical
performance. In other cases, however, long-time stability is almost
impossible to achieve for some methods.

A common strategy to develop stable numerical methods is to request
that these obey a number of exact features (although this does not
ensure the stability or the performance). There are a series of exact
conditions that can be easily derived. For example, it is well known
that, for linear systems with Hermitian Hamiltonians, the propagator
is unitary
\begin{equation}
  \label{eq:unitary}
  \hat{U}^\dagger(t, t-\Delta t) = \hat{U}^{-1}(t, t-\Delta t)
\end{equation}
This property ensures that the KS wave-functions remain orthonormal
during the time-evolution. Algorithms that severely
violate~\eqref{eq:unitary} will have to orthogonalize regularly the
wave-functions, a rather expensive ($N^3$) operation, especially for
large systems. Note that for the TDKS equationss it is not, strictly speaking, correct to
speak of unitarity due to the nonlinear character of the propagators even if
the orthogonality condition still holds among the KS orbitals (see the
discussion in Ref.~\citenum{Andrade2009}).

For systems that do not contain a magnetic-field or a spin-orbit
coupling term (or any other term that breaks time-reversal symmetry),
the evolution operator fullfils
\begin{equation}
  \hat{U}(t,t-\Delta t) = \hat{U}^{-1}(t-\Delta t,t)
  \,.
\end{equation}
This relation is rather important in order to ensure stability of the
numerical propagator, and it is often violated by many explicit
methods.

\subsection{Symplecticity}


The \emph{geometrical structure} of an ODE system, as well as that of
its numerical representation (i.e. the propagator), is another
important issue to consider.\cite{Hairer2006} In this context, an
important property is symplecticity. A differentiable map $g: \mathbb{R}^{2n} \to
\mathbb{R}^{2n}$ is symplectic if and only if
\begin{equation}
    \frac{\partial g}{\partial y}^T J \frac{\partial g}{\partial y} = J\,,
    \qquad
    J = \left[\begin{array}{cc}
        0 & I
        \\
        -I & 0
      \end{array}\right]\,.
\end{equation}
Given any system of ODEs, the \emph{flow} is a differentiable
map. The first requirement for a flow to be symplectic is that the
system is formed by an even number of real equations.  Any
\emph{complex} system, however, may be split into its real and
imaginary parts, and is equivalent to a system with
an even number of real equations.

The system of equations is also required to be \emph{Hamiltonian}: a
system is Hamiltonian if it follows the equation of motion
\begin{equation}
  \dot{y} = J^{-1}\nabla H(y)\,,
\end{equation}
where $y \in \mathbb{R}^{2n}$, and $H$ is some scalar function of
$y$. It is usual to decompose $y=(q, p)^\text{T}$, leading to the
well-known Hamilton equations of motion
\begin{subequations}
\begin{align}
  \displaystyle \dot{q_i} & = \phantom{-}\frac{\partial H(p,q)}{\partial p_i}
  \\
  \displaystyle \dot{p_i} & = -\frac{\partial H(p,q)}{\partial q_i}
\end{align}
\end{subequations}
with $q_i$ and $p_i$ elements of the vectors $q$ and $p$. The flow of
a Hamiltonian system is symplectic. Roughly speaking, the inverse is
also true.\cite{Poincare1999,Hairer2006}

One can easily prove that the (usual) Schr\"odinger equation
\begin{subequations}
\begin{align}
  \label{eq:sch1}
  i\frac{\rm d}{{\rm d}t}\vert\Psi(t)\rangle & =  \hat{H}\vert\Psi(t)\rangle\,,
  \\
  \vert\Psi(0)\rangle & =  \vert\Psi_0\rangle\,.
\end{align}
\end{subequations}
forms a Hamiltonian system,\cite{Heslot1985} and is therefore symplectic. It is possible
to perform the derivation in coordinate space, but the proof is
somewhat simpler if we expand the wave-function in a given basis set
\begin{equation}
  \vert\Psi(t)\rangle = \sum_i c_i(t) \vert\Psi_i\rangle
\end{equation}
where $\lbrace\vert\Psi_i\rangle\rbrace$ forms an orthonormal basis
and $c_i(t)=\langle\Psi_i\vert\Psi(t)\rangle$ are the time-dependent
expansion coefficients. The Schr{\"{o}}dinger equation is thus
transformed into:
\begin{subequations}
\begin{align}
  \label{eq:sch2}
  \dot{c} & = -i H{c}\,,
  \\
  c_i(0) & = \langle\Psi_i\vert\Psi_0\rangle\,
\end{align}
\end{subequations}
where the Hamiltonian matrix $H$ defined by $H_{ij} =
\langle\Psi_i\vert{H}\vert\Psi_j\rangle$, and $c$ is the vector of the
coefficients. We now split the coefficients $c_i$ of the wave-function
into their real and imaginary parts
\begin{equation}
  c_i = \tfrac{1}{\sqrt{2}}(q_i + ip_i)\,,
\end{equation}
i.e., $q_i = \sqrt{2}\:\Re c_i$, $p_i = \sqrt{2}\:\Im c_i$.  We can
now define a Hamiltonian function of the vectors $q$ and $p$
\begin{align}
  \nonumber
  H(q, p) & = \langle \Psi(q,p) \vert \hat{H} \vert \Psi(q,p)\rangle
  \\\nonumber
  & = \tfrac{1}{\sqrt{2}}(q -i p)^\text{T} (\Re H + i \Im H) \tfrac{1}{\sqrt{2}}(q +i p)
  \\\nonumber
  & = \tfrac{1}{2}q^\text{T}\Re H q + \tfrac{1}{2}p^\text{T}\Re H p + p^\text{T}\Im H q
  \,,
\end{align}
where we have used the fact that, if $H$ is Hermitian, then $H(q,p)$
must be real, and the real part of $H$ is symmetric ($\Re H^\text{T} =
\Re H$) while its imaginary part is anti-symmetric ($\Im H^\text{T} =
-\Im H$).  We can now calculate the partial derivatives
\begin{subequations}
\label{eq:partialHqp}
\begin{align}
  \frac{\partial H(q, p)}{\partial q_i} & =  \sum_j\left(\Re H_{ij} q_j - \Im H_{ij} p_j\right)\,,
  \\
  \frac{\partial H(q, p)}{\partial p_i} & =  \sum_j\left(\Re H_{ij} p_j + \Im H_{ij} q_j\right)\,.
\end{align}
\end{subequations}

In order to find the equations of motion for the ${\bf q}$ and ${\bf
  p}$ coordinates, we rewrite Eq.~\eqref{eq:sch2} as:
\begin{equation}
  \label{eq:qdotpdot}
  \dot{q_i} + i\dot{p_i} = -i \sum_j (\Re H_{ij} + i \Im H_{ij}) (q_j + i p_j)\,,
\end{equation}
The proof follows by separating the real and imaginary parts of
\eqref{eq:qdotpdot} and comparing them to Eqs.~\eqref{eq:partialHqp}. The
Schr{\"{o}}dinger's equation forms a Hamiltonian, symplectic system.

Whether or not an ODE system is symplectic has important theoretical
consequences (the flow preserves the volume in phase space, for
example). Numerically, the algorithm that we choose to approximate the
real flow defines a \emph{numerical flow} that may or may not be
symplectic. It is of course convenient for it to be: for example, one
can demonstrate~\cite{Benettin1994} that symplectic numerical flows
lead to long term stability of the energy, that typically oscillates
around its true value. Usually, the error in the conservation of other
constants of motion also behaves better when symplectic algorithms are
used. In the following, we shall prove, following a similar procedure
to the one above for the Schr{\"{o}}dinger equation, that the TDKS
equations, in the {\it adiabatic} approximation, form a symplectic,
Hamiltonian system. Therefore, it is convenient (although not strictly
necessary) to choose symplectic algorithms to approximate the TDKS
propagator.

\subsection{Symplecticity and the TDKS equations}

For the TDKS equations, Eqs.~\eqref{eq:KSE}, the Hamiltonian operator can be written as
\begin{equation}
  \hat{H}[n(t)] = 
  \hat{T} + \hat{V} + \hat{V}_{\rm Hxc}[n(t)]\,,
\end{equation}
where the terms represent the kinetic energy operator, the external
potential, and the Hartree-exchange-correlation (Hxc) potential. In
the coordinate representation, we have
\begin{multline}
  \langle {\bf r}\sigma\vert\hat{H}[n(t)]\vert\varphi_m(t)\rangle  =
  -\tfrac{1}{2}\nabla^2 \varphi_m({\bf r}\sigma,t)
  \\ 
  + v({\bf r}) \varphi_m({\bf r}\sigma,t) 
  + v_{\rm Hxc}[n(t)]({\bf r}) \varphi_m({\bf r}\sigma,t)\,.
\end{multline}
We now expand the KS orbitals in a one-electron basis
$\lbrace\vert\phi_i\rangle\rbrace$
\begin{equation}
  \vert\varphi_m\rangle = \sum_i c_{mi}\vert\phi_i\rangle\,.
\end{equation}
The TDKS equations are thus transformed into the initial value problem
\begin{subequations}
\begin{align}
  \dot{c}_m & = -i H[c]c_m
  \\
  c_{mi}(0) & = \langle \phi_i\vert\varphi_m^0\rangle\,,
\end{align}
\end{subequations}
where the matrix $H[c]$ is given by
\begin{equation}
  H[c]_{ij} = \langle\phi_i\vert\hat{H}[c]\vert\phi_j\rangle\,.
\end{equation}
Note that the dependence on the (instantaneous) density is rewritten
as a dependence on the full set of coefficients $c$. We again split the
coefficients into their real and imaginary parts
\begin{equation}
  c_{mi} = \tfrac{1}{\sqrt{2}}(q_{mi} + ip_{mi})\,.
\end{equation}
The TDKS equation may then be rewritten as
\begin{equation}
  \dot{q}_m + i\dot{p}_m = -i (\Re H[q,p] + \Im H[q,p])(q_m + ip_m)
\end{equation}
and separating into real and imaginary parts
\begin{subequations}
\label{eq:TDKSrealimag}
\begin{align}
  \dot{q}_m & = \phantom{-}\Im H[q,p]q_m + \Re H[q,p]p_m\,.
  \\
  \dot{p}_m & = -\Re H[q,p]q_m + \Im H[q,p]p_m\,.
\end{align}
\end{subequations}
In order to rewrite the TDKS system as a classical Hamiltonian system,
we need to find a Hamiltonian function $H(q,p)$. It can be easily seen
that the non-interacting energy of the KS system does not
work. However, we can use the \emph{ground-state} energy functional,
which is given by
\begin{equation}
  E[n] = T_\text{S}[n] + V[n] + E_{\rm Hxc}[n]\,,
\end{equation}
evaluated adiabatically with the {\it time-dependent}
density. Remembering that the density is evaluated from the KS
orbitals, we can write the energy as a functional of these
\begin{equation}
  E[\varphi] = T_\text{S}[\varphi] + V[\varphi] + E_{\rm Hxc}[\varphi]\,.
\end{equation}
Representing the orbitals by the new variables $(q,p)$, we define a
Hamiltonian function
\begin{equation}
  H(q,p) = T_\text{S}(q,p) + V(q,p) + E_{\rm Hxc}(q,p)\,.
\end{equation}
The first two terms can be treated exactly in the same way as for the
standard Schr{\"o}dinger equation. The non-interacting kinetic energy
function reads
\begin{multline}
  T(q,p) = \tfrac{1}{2}\sum_m q_m\Re T_\text{S} q_m
  +  \tfrac{1}{2}\sum_m p_m \Re T_\text{S} p_m
  \\
  + \sum_m p_m \Im T_\text{S} q_m\,,
\end{multline}
while the external potential is
\begin{multline}
  V(q,p) = \tfrac{1}{2}\sum_m q_m\Re V q_m +  \tfrac{1}{2}\sum_m p_m \Re V p_m
  \\
  + \sum_m p_m \Im V q_m
  \,.
\end{multline}
Calculating the partial derivatives of the previous expressions, we
arrive at
\begin{subequations}
\begin{align}
  \frac{\partial T(q,p)}{\partial q_{mi}} & = \sum_j \left(\Re T_{ij} q_{mj}
  - \Im T_{ij} p_{mj}\right)\,,
  \\
  \frac{\partial T(q,p)}{\partial p_{mi}} & = \sum_j \left(\Re T_{ij} p_{mj}
  + \Im T_{ij} q_{mi}\right)\,,
\end{align}
\end{subequations}
and with a similar expression for ${\partial V(q,p)}/{\partial
  q_{mi}}$ and ${\partial V(q,p)}/{\partial p_{mi}}$. Using
\eqref{eq:TDKSrealimag}, we see that these two terms verify 
the necessary conditions for a Hamiltonian system. We are left with
the term $E_{\rm Hxc}(q,p)$. Its partial derivatives can be computed
with the help of the chain rule
\begin{subequations}
\label{eq:dehxcdqp}
\begin{align}
  \label{eq:dehxcdq}
  \frac{\partial E_{\rm Hxc}(p,q)}{\partial q_{mi}} & = \int\!\!{\rm d}^3r\;
  \frac{\delta E_{\rm Hxc}}{\delta n(q,p; {\bf r})}
  \frac{\partial n(q,p; {\bf r})}{\partial q_{mi}}
  \\
  \label{eq:dehxcdp}
  \frac{\partial E_{\rm Hxc}(p,q)}{\partial p_{mi}} & = \int\!\!{\rm d}^3r\;
  \frac{\delta E_{\rm Hxc}}{\delta n(q,p; {\bf r})}
  \frac{\partial n(q,p; {\bf r})}{\partial p_{mi}}
  \,.
\end{align}
\end{subequations}
The density $n(q,p, {\bf r})$ is the one that corresponds to the set
of Kohn-Sham orbitals defined by the $(q,p)$ coordinates. The
functional derivative of $E_{\rm Hxc}$ is the Hartree, exchange, and
correlation potential
\begin{equation}
  \frac{\delta E_{\rm Hxc}(q,p)}{\delta n(q,p; {\bf r})} = 
  v_{\rm Hxc}(q,p; {\bf r})\,.
\end{equation}
In order to compute the partial derivatives of the density with
respect to $q$ and $p$, one needs to write it in terms of those
variables
\begin{equation}
  n(q,p; {\bf r}) = \tfrac{1}{2}\sum_{\sigma, m \atop ij} (q^m_i -i
  p^m_i)(q^m_j+ip^m_j) \phi_i^*({\bf r}\sigma)\phi_j({\bf r}\sigma)\,.
\end{equation}
Then
\begin{subequations}
\begin{multline}
  \frac{\partial n(q,p; {\bf r})}{\partial q_{mi}} =
    \sum_j q_{mj} \Re \sum_\sigma \phi_i^*({\bf r}\sigma) \phi_j({\bf r}\sigma)
    \\
    - \sum_j p_{mj} \Im \sum_\sigma \phi_i^*({\bf r}\sigma) \phi_j({\bf r}\sigma)
    \,,
\end{multline}
and
\begin{multline}
  \frac{\partial n(q,p; {\bf r})}{\partial p_{mi}} =
  \sum_j q_{mj} \Im \sum_\sigma \phi_i^*({\bf r}\sigma) \phi_j({\bf r}\sigma)
  \\
  + \sum_j p_{mj} \Re \sum_\sigma \phi_i^*({\bf r}\sigma) \phi_j({\bf r}\sigma)
\end{multline}
\end{subequations}
Plugging these expressions into Eqs.~\eqref{eq:dehxcdqp} results in
\begin{subequations}
\begin{align}
  \frac{\partial E_{\rm Hxc}(p,q)}{\partial q_{mi}} &=
  \sum_j \left(\Re V^{\rm Hxc}[q,p]_{ij}q_{mj} - \Im V^{\rm Hxc}[q,p]_{ij}p_{mj}\right)
  \\
  \frac{\partial E_{\rm Hxc}(p,q)}{\partial p_{mi}} &=
  \sum_j \left(\Im V^{\rm Hxc}[q,p]_{i,j}q_{mj} + \Re V^{\rm Hxc}[q,p]_{ij}p_{mj}\right)
\end{align}
\end{subequations}
where the matrix $V^{\rm Hxc}[q,p]$ is given by:
\begin{equation}
  V^{\rm Hxc}[q,p]_{ij} = \langle\phi_i \vert\hat{V}_{\rm Hxc}[q,p] \vert \phi_j\rangle\,.
\end{equation}
Therefore, the partial derivatives of $E_{\rm Hxc}$ also have the
right structure, which concludes the proof that the TDKS equations
form a Hamiltonian system.

\section{Results}
\label{sec:results}
In order to analyze the performance of the integration schemes we used
a ``real world'' benchmark based on the propagation of a benzene
molecule. We placed the molecule in a
spherical simulation box of radius $r=12$~a.u., with a grid spacing of
$a=0.4$~a.u. At time zero, the system is subject to an instantaneous
perturbation:
\begin{equation}
  \varphi_j^{\rm GS} \rightarrow \varphi_j(t=0^+) = e^{ikz}\varphi_j^{\rm GS}\,,
\end{equation}
i.e. each KS orbital, initially at its ground-state equilibrium value
$\varphi_j^{\rm GS}$ is transformed at time zero into a slightly
perturbed orbital $\varphi_j(t=0^+)$, corresponding to a sudden
application of an electric field with strength $k=0.1$ a.u. in the
$z$-direction.  Then it evolves freely for a total propagation time
$T=2\pi$~a.u.  We compared both the wave-function and the energy
obtained at the end of the run with a reference ``exact'' calculation,
performed with a very small time step and the explicit RK4
propagator. The error in the wave-function is then defined as
\begin{equation}
  E_\textrm{wf}(T,\Delta t) = \sqrt{\sum_m \vert\vert \varphi_m(T) - \varphi_m^\textrm{exact}(T)\vert\vert^2}\,,
\end{equation}
and the error in the energy is defined as
\begin{equation}
  E_\textrm{energy}(T,\Delta t) = \vert E(T)-E^\textrm{exact}(T)\vert\,,
\end{equation}
where $\varphi_m^\textrm{exact}$ and $E^\textrm{exact}$ are the KS
orbitals and the energy obtained from the ``exact'' calculation.

\subsection{Exponential midpoint rule}

We used the exponential midpoint rule (EMR), one of the
propagators studied in Ref.~\citenum{Castro2004}, as a base for comparison with the new schemes.
The EMR prescribes:
\begin{equation}
  \label{eq:emr}
  \varphi(t) = \exp\left( -i \Delta t \hat{H}[\overline{\varphi}](t-\Delta t/2)\right)
  \varphi(t-\Delta t)\,
\end{equation}
where $\overline{\varphi}$ is the average wavefunction:
\begin{equation}
  \label{eq:emr2}
  \overline{\varphi} = \tfrac{1}{2}[\varphi(t)+\varphi(t-\Delta t)]\,.
\end{equation}
The EMR is second order in $\Delta t$, symplectic, and preserves
time-reversal symmetry. It is also an implicit scheme as it requires the
Hamiltonian calculated with the average wave-function. The non-linear
equations~\eqref{eq:emr} and \eqref{eq:emr2} can be solved, e.g., by
iteration until self-consistence is achieved. The first iteration can
be started by making use of an extrapolated Hamiltonian
\begin{equation}
  \label{eq:aemr}
  \varphi^{(1)}(t) = \exp\left( -i \Delta t \hat{H}^{\rm extr}_{(t-\Delta t/2)} \right)
  \varphi(t-\Delta t)\,
\end{equation}
We will use the shorthand notation $\hat{H}^{\rm extr}_{(\tau)}$ for a
Hamiltonian that is obtained via extrapolation or interpolation to
time $\tau$ from a number $p$ of known Hamiltonians:
$\hat{H}_{t-\Delta t}, \hat{H}_{t-2\Delta t},\dots,\hat{H}_{t-p\Delta
  t}$. We will also use the notation
\begin{equation}
  \hat{H}_{(\tau)} = \hat{H}[\varphi(\tau)](\tau)\,.
\end{equation}

In practice, most of the times one does not iterate the
self-consistent procedure, but uses Eq.~\eqref{eq:aemr} directly. This
leads to an \emph{explicit} EMR, that is the method used in the remainder
of this work. Of course, this approximated method no longer fulfills
the exact properties stated above.

The definition of the algorithm must be complemented with a recipe to
compute the action of the exponential of an operator on a
vector. There are a variety of possibilities, passing by a
truncated Taylor expansion, a Lanczos expansion, the split-operator
scheme (as well as any of the higher-order variants of this), etc. For
our purposes we decided to use the first, namely a Taylor
expansion truncated to fourth order.

One may also design other exponential-based methods that can be
considered variations of the EMR. For example, in
Ref.~\citenum{Castro2004} we defined the ``enforced time-reversal
symmetry'' (ETRS) scheme
\begin{multline}
  \label{eq:etrs}
  \varphi(t) = 
  \exp\left( -i \frac{\Delta t}{2} \hat{H}_{(t)}\right)\times \\
  \exp\left( -i \frac{\Delta t}{2} \hat{H}_{(t-\Delta t)}\right)
  \varphi(t-\Delta t)\,.
\end{multline}
This algorithm was designed to improve on the preservation of
time-reversal symmetry. It is also an implicit method, and the
non-linear equation~\eqref{eq:etrs} can be solved
iteratively. Alternatively, one can use an extrapolated Hamiltonian
$\hat{H}^\text{extr}_{(t)}$ in Eq.~\eqref{eq:etrs}, leading to the
approximate ETRS (AETRS) algorithm.

\subsection{Commutator-free Magnus expansions}

\begin{figure}
  \centerline{
    \includegraphics[width=\columnwidth]{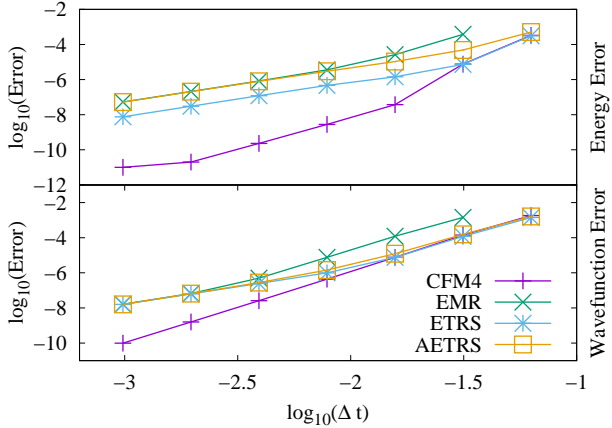}
  }
\caption{
  \label{fig:common1}
  Error in the total energy (top panel) and in the wave-function (bottom
  panel), as a function of the time-step, for the various 
  reference propagators (ETRS, AETRS and EMR) and for the CFM4
  propagator.}
\end{figure}

\begin{figure}
  \centerline{ \includegraphics[width=1.0\columnwidth]{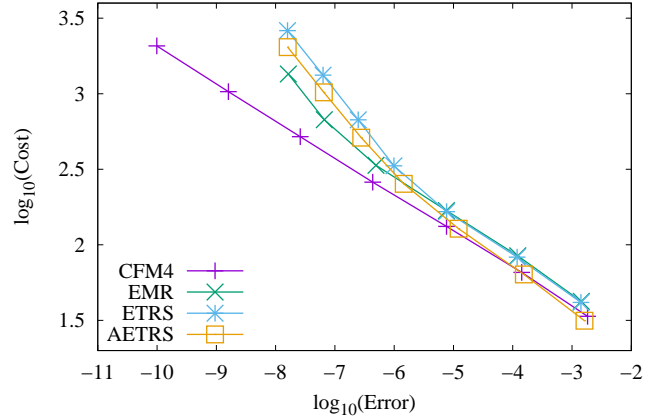} }
  \caption{
    \label{fig:common2}
    Cost of the method, as a function of the error obtained (in the
    wave-function), for the various reference
    propagators (ETRS, AETRS and EMR) and for the CFM4 propagator.  }
\end{figure}

Restricting the discussion momentarily to linear systems, for 
time-dependent Hamiltonians,
the evolution operator~\eqref{eq:toevolution} has a rather complicated
form involving an integral over time and a time-ordering operator.  It
is natural to wonder, however, if there exists an operator
$\hat{\Omega}(t,t-\Delta t)$ that makes the following expression exact
\begin{equation}
  \hat{U}(t,t-\Delta t) = \exp\left( \hat{\Omega}(t,t-\Delta t) \right)\,.
\end{equation}
In 1954 Magnus~\cite{Magnus1954} showed that, for some neighborhood of
$t$, there exists an infinite series such that
\begin{equation}
  \hat{\Omega}(t,t-\Delta t)=\sum\limits^{\infty}_{m=1}\hat{\Omega}_{m}(t,t-\Delta t)\,,
\end{equation}
and provided a recursive relation to find the operators
$\hat{\Omega}_{m}$. This recursive relation involves nested
commutators of the Hamiltonian at different times.  To obtain a
Magnus propagator of order $2n$, $\hat{U}_{M(2n)}$, one truncates the
series at the $n$-th term, and approximates each term 
with some $n$-th order quadrature formula.  As an example
of this procedure, the aforementioned EMR can be regarded as a Magnus
expansion of order two (although, strictly speaking, only for linear systems):
\begin{equation}
  \hat{U}_{\rm EMR}(t,t-\Delta t)= \exp\left( -i\Delta
  t\hat{H}_{(t-\Delta t/2)}\right) = \hat{U}_{M(2)}\,.
\end{equation}
This second order formula is unusual as it does not involve
commutators. For higher orders, the main difficulty arises from the
evaluation of the commutators. To circumvent this problem, Blanes et
al.~\cite{Blanes2006} developed a series of alternative Magnus
expansions that are free of commutators, and also address the
nonlinear case.

We implemented the fourth order commutator-free version of the
Magnus expansion, presented in Eq.~(43) of
Ref.~\citenum{Blanes2006}, and labelled as the ``Method 4'' in
page 6 of Ref.~\citenum{Auer2017}. This method, that we will refer
to in the following by ``CFM4'', is given by:
\begin{multline}
  \varphi(t) = 
  \exp\left( -i \Delta t \alpha_1 \hat{H}_{(t_1)}
  -i \Delta t \alpha_2 \hat{H}_{(t_2)} \right) \times
  \\
  \exp\left( -i \Delta t \alpha_2 \hat{H}_{(t_1)}
  -i \Delta t \alpha_1 \hat{H}_{(t_2)} \right)
  \varphi(t-\Delta t)\,.
\end{multline}
for some carefully chosen constants $\alpha_1,\alpha_2$ and intermediate times $t_1,t_2$.
The application of this method to the nonlinear TDKS equations leads
again to an implicit rule, as we need $\hat{H}_{(t_1)}$ and
$\hat{H}_{(t_2)}$.  Therefore, we have implemented an approximate
version, again relying on extrapolated Hamiltonians. If this
extrapolation is peformed at fourth order (i.e. using at least four
previous steps), then the order of the method is preserved.

Figures.~\ref{fig:common1} and~\ref{fig:common2} depict the results
obtained with the CFM4, EMR, ETRS and AETRS methods.  The top (botton)
panel of Fig.~\ref{fig:common1} shows the error in the energy
(wave-function) as a function of the time-step. We used logarithmic
scales in both axes, so that the curves become straight lines in the
small $\Delta t$ limit (until numerical precision is reached). The
slope of those lines is given by the order of each method -- at least
for the error in the wave-function.  For larger values of the time-step, the
curves are no longer straight lines, and may actually exhibit a faster
behavior: for example, the EMR, ETRS and AETRS methods behave as
fourth order propagators for larger $\Delta t$, whereas their order is
actually two.  As we can see in Fig.~\ref{fig:common1}, for the
largest time-steps (up to $10^{-2}$ a.u.) all the methods have
similar precision, except the EMR, which becomes unstable (this is the reason
why this data point is missing). When the
time-step decreases, EMR, ETRS and AETRS behave as order-two methods
while CFM4 maintains its fourth order throughout the whole range of $\Delta
t$. This makes CFM4 significantly more precise than the other
propagators for $\Delta t<10^{-2}$~a.u.

In Fig.~\ref{fig:common2} we show the cost (measured in seconds) of
the propagation as a function of the error in the wave-function, again
in logarithmic scale. From these kind of plots one can identify the
best performing method for a given \emph{required} precision. This
required precision must be decided a priori by the user, and it is
problem dependent.  For the largest values of the error the
performance of all the integrators is very similar.
For smaller values, EMR, ETRS and AETRS have similar cost, but
CFM4 is significantly faster. This makes CFM4 the best method
overall.

\subsection{Multistep methods}
\label{subsec:multistep}

In 1883 J. C. Adams and F. Bashforth proposed multistep methods in the
context of fluid mechanics.\cite{Adams1883} These methods
use $s>1$ previous steps in order to calculate the following one. 
They require a starting procedure to provide
those first $s$ steps. The simplest procedure consists in using
a single-step method. In our case we used the standard explicit
fourth-order RK (described below).

We examined linear multistep formulas given by
\begin{multline}
  \varphi(t)+\sum\limits^{s}_{k=1}a_{s-k}\varphi(t-k\Delta t)= \\
  \Delta t\sum\limits^{s}_{k=0}b_{s-k}f(t-k\Delta t,\varphi(t-k\Delta t))\,,
\end{multline}
where $\{a_{k}\}_{k=0}^{s-1}$ and $\{b_{k}\}_{k=0}^{s}$ are the coefficients
that determine the method.  If $b_s = 0$, then the method is explicit,
since the equation is an explicit formula for $\varphi(t)$.  If $b_s
\ne 0$ then the method is implicit, as it provides a relation between
$\varphi(t)$ and $f(\varphi(t),t)$. If we consider the 
dynamical function relevant for TDDFT
\begin{equation}
  f(t,\varphi)=-iH_{(t)}\varphi\,,
\end{equation}
and we define the shorthand notation
\begin{equation}
  \varphi^{(k)} = H_{(t-k\Delta t)}\varphi(t-k\Delta t)
\end{equation}
we finally arrive at
\begin{multline}
  (I+b_{s}i\Delta tH_{(t)})\varphi(t)= \\
  -\sum\limits^{s}_{k=1}[a_{s-k}\varphi(t-k\Delta t)+b_{s-k}i\Delta t \varphi^{(k)}].
\end{multline}

The first multistep integrators that we studied belong to the family
of explicit Adams methods, also known as Adams-Bashforth (AB)
methods. They are explicit ($b_s=0$) and the coefficients $a$ are:
$a_{s-1} = -1$ , and $a_{s-2}=\dots=a_{0}=0$.  The remaining $b_{k}$
are chosen such that the methods have order $s$, which determines them
uniquely.  The method then reads:
\begin{equation}
  \varphi(t) = \varphi(t-\Delta t)
  - \sum_{k=1}^s b^{\rm AB}_{s-k} i\Delta t \varphi^{(k)}\,.
\end{equation}

The implicit Adams, or Adams-Moulton (AM) family is similar to the
Adams-Bashforth methods in that they also have $a_{{s-1}}=-1$ and
$a_{s-2}=. . . =a_{0}=0$:
\begin{multline}
  \label{eq:am}
  (I+b^{{\rm AM}}_{s}i\Delta tH_{(t)}) \varphi(t)= \\
  \varphi(t-\Delta t) 
  -\sum\limits^{s}_{k=1}b^{\rm AM}_{s-k}i\Delta t \varphi^{(k)}\,.
\end{multline}
Again, the $b$ coefficients are chosen to obtain the highest possible
order.  The Adams-Moulton methods are implicit methods, since the
restriction $b_s = 0$ is removed. This fact permits the increase of
the order of the error: an $s$-step Adams-Moulton method is of order
$s+1$, while an $s$-step Adams-Bashforth method is only of order $s$.

Equation~\eqref{eq:am} was solved iteratively. We also implemented a
``linearized'' version of the Adams-Moulton formula (lAM), where we
used an extrapolation of the Hamiltonian at time $t$, thereby
transforming Eq.~\eqref{eq:am} into a linear equation.  Another
possible simplification of the Adams-Moulton formula regards the use
of the so-called ``predictor-corrector'' schemes, which avoid the
linear system solution altogether by turning the implicit method into
an explicit one. In our implementation, it consists of using
Adams-Bashforth to get an approximated (``predictor'')
$\tilde{\varphi}(t)$, and use this to obtain the Hamiltonian in the
left-hand side of Eq.~\eqref{eq:am}. We named this procedure the
Adams-Bashforth-Moulton (ABM) method.

The backward differentiation formulas (BDF) are implicit methods with
$b_{s-1} = \dots = b_0 = 0$ and the other coefficients chosen such
that the method has order $s$ (the maximum possible). These methods
are especially suited for the solution of stiff differential
equations. The general formula for a BDF can be written as:
\begin{equation}
  (I+b^{{\rm BDF}}_{s}i\Delta tH_{(t)})\varphi(t)=-\sum\limits^{s}_{k=1}a^{{\rm BDF}}_{s-k}\varphi(t-k\Delta t).
\end{equation}

\begin{figure}
  \centerline{
    \includegraphics[width=\columnwidth]{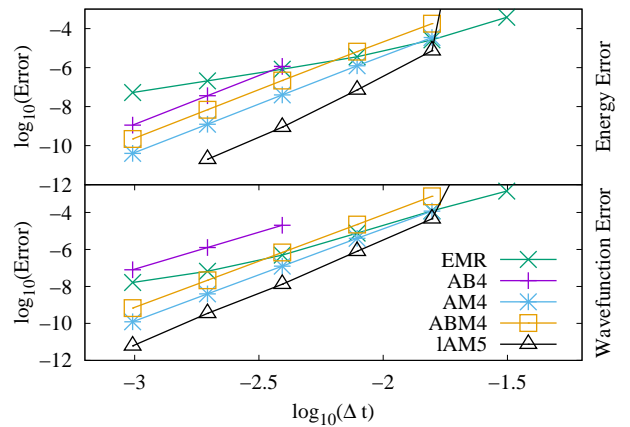}
  }
  \caption{
    \label{fig:multistep1}
    Error in the total energy (top panel) and in the wave-function
    (bottom panel), as a function of the time-step, for the various
    multistep methods (AB, AM, ABM and linearized AM) and for the EMR
    propagator.}
\end{figure}

\begin{figure}
  \centerline{
    \includegraphics[width=0.8\columnwidth]{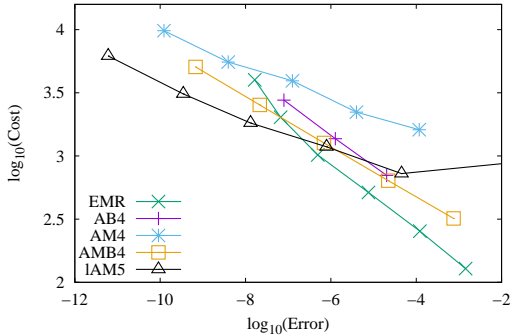}
  }
  \caption{
    \label{fig:multistep2}
    Cost of the method, as a function of the error obtained (in the
    wave-function), for the various multistep methods (AB, AM, ABM and
    linearized AM) and for the EMR propagator.  }
\end{figure}

We implemented these families of integrators in octopus, ran these
five methods with steps $s=1,\dots,5$, and compared them among each
other and with the EMR. Figures.~\ref{fig:multistep1}
and~\ref{fig:multistep2} show the best candidate from each family.
The number accompanying the name of the propagator indicates the
number of previous steps $s$ used in the calculation.  As we can see,
the EMR is more stable than any of the multistep methods for large
time-steps (especially AB4, which is the most unstable), but is
outclassed in precision by every other propagator. This is not
surprising, as they are methods of order 4 (AB4 and ABM4), 5 (AM4) or
6 (linearized AM5). The most precise method for a given time-step is
the lAM5, reaching the numerical precision of our machines for the
smallest time-steps.

In Fig.~\ref{fig:multistep2} we plot the cost of the methods as a function
of the error. AB4 cannot compete in precision, stability or
performance with the EMR.  For error values larger than $10^{-7}$ the
EMR is the fastest propagator, while for smaller values it is overcome
by AMB4 and linearized AM5. AM4 is, as expected, the most
computationally expensive method, with the linearization procedure
dramatically improving its speed.

\begin{figure}
  \centerline{
    \includegraphics[width=\columnwidth]{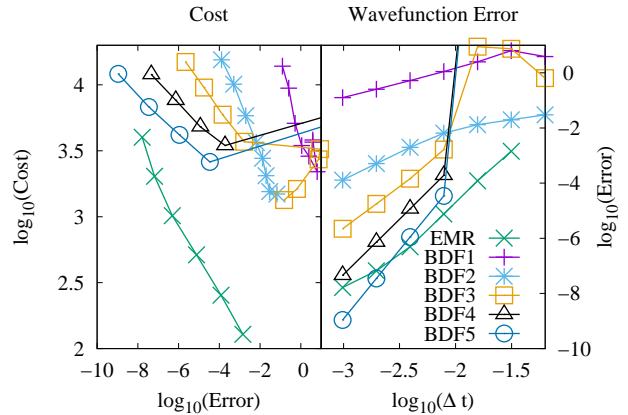}
  }
  \caption{
    \label{fig:multistep3}
    Left: Cost as a function of the error for the BDF methods (going
    from $s=1$ to 5) and for the EMR propagator.  Right: Error in the
    propagated wave-function, as a function of the time-step, for the
    BDF methods (going from 1 to 5 previous steps) and for the EMR
    propagator.}
\end{figure}

In Fig.~\ref{fig:multistep3} we represent the BDF results for
$s=1,\dots,5$. Our aim here is to illustrate one important
characteristic of the multistep methods, namely that the cost does not
increase significantly with the number of previous steps $s$.  This
can be clearly seen on the left panel of
Fig.~\ref{fig:multistep3}. Furthermore, in the right panel we can see
that each extra step included in the method increases its order by
one. Then, why not increasing the number of steps to a very larger
number? First, there is a memory issue, as the previous $s$ steps have
to be stored in memory.  But more importantly, as the number of
previous steps increases, the stability of the method decreases. This
can be seen in both panels of Fig~\ref{fig:multistep3}. Both BDF1 and
BDF2 have better stability properties than EMR, but as soon as we make
$s\geq 3$ we need to reduce the time-step by a factor of 16 to avoid
the breaking down of the method. This reduction of the stability
region with the number of steps seems to hold for all linear multistep
methods. Moreover, for BDF there is a mathematical proof that states
that for $s\geq 7$ these methods are unstable (check section III.3
from Ref.~\citenum{Hairer1996} for a more detailed explanation).

Finally, one important caveat of multistep methods is that they cannot be
symplectic. In fact, the definition does not even apply, as a multistep
algorithm is a map from several previous steps into the next one, and one
cannot speak of a flow in the usual way. There are however some ways to understand
symplecticity also for these methods,\cite{Hairer2006} but the conclusion is
in any case negative, and the long-term stability properties of these methods
is disappointing.

\subsection{Runge-Kutta schemes}
\label{subsec:rk}

\subsubsection{``Standard'' Runge-Kutta schemes}

The Runge-Kutta (RK) schemes form a family of methods developed around
1900 by C. Runge and M. W. Kutta.\cite{Butcher1996} Let $b_i$,
$a_{ij}$ ($i, j = 1, \dots, s$) be real numbers and
$c_i=\sum^{i-1}_{j=1}a_{ij}$.  The scheme
\begin{equation}
  \varphi(t) = \varphi(t-\Delta t) +\Delta t \sum\limits^{s}_{i=1}b_{i}Y_{i}\,,
\end{equation}
where the functions $Y_{i}$ are defined as
\begin{equation}
  Y_{i} = f(\varphi(t-\Delta t) + \Delta t\sum_{j=1}^s a_{ij}Y_j, t_{i})\,,
\end{equation}
at the time-steps
\begin{equation}
  t_{i} = t-\Delta t+c_i\Delta t\,,\qquad i = 1, . . . , s\,,
\end{equation}
is called an $s$-stage RK scheme. To specify a particular method, one
needs to provide the integer $s$ (the number of stages), and the
coefficients $a_{ij}$, $b_i$, and $c_i$ (for $i = 1, 2, \dots,
s$). These are usually arranged in a mnemonic device, known as a
Butcher tableau:
\begin{equation}
\centerline{
\begin{tabular}{ p{1cm} | p{1cm}  p{1cm}  p{1cm}  p{1cm} }
  $c_1$ & $a_{11}$ & $a_{12}$ & \dots & $a_{1s}$ \\
  $c_2$ & $a_{21}$ & $a_{22}$ & \dots & $a_{2s}$ \\
  $\vdots$ & $\vdots$ & $\vdots$ & $\ddots$ & $\vdots$ \\
  $c_s$ & $a_{s1}$ & $a_{s2}$ & $\dots$ & $a_{ss}$ \\
  \hline
       & $b_1$ & $b_2$ & $\dots$ & $b_s$
\end{tabular}
}
\end{equation}

When $a_{ij}=0$ for $i\leq j$ the method is explicit, whereas in all
other cases the method is implicit. Explicit RK methods are generally
unsuitable for the solution of stiff equations because their region of
absolute stability is small. These shortcomings motivated the
development of implicit methods. They are visually easy to identify
looking at their tableaux, as they include non-zero entries in the upper
triangle.

For the implicit methods we need to solve a system of algebraic
equations, the dimension of which grows with the number of
stages: For a method with $s$ stages, the equation has $m\times s$
unknowns, where $m$ is the dimension of the original system. In
contrast, linear multistep methods only require the solution of
$m$-dimensional algebraic equations.

For a RK scheme to be symplectic, one can prove\cite{Sanz-Serna1988} that the $s\times
s$-matrix $M$ with coefficients
\begin{equation}
  \label{eq:symRK}
  m_{ij} = b_{i}a_{ij}+b_{j}a_{ji}-b_{i}b_{j}
\end{equation}
has to satisfy $M=0$. This implies that no explicit RK scheme can be symplectic.

We studied the RK propagators up to order four. The reason behind this choice
is that, up to this order, the required number of stages $s$ for explicit
methods is equal to the desired order of the method. From order five onward,
however, $s$ is strictly greater than the desired
order.\cite{Butcher1987,Butcher1996,Hairer2006} Therefore, the precision
gained by increasing the order does not compensate for the increase in the
computational cost.  Regarding explicit methods, the most widely known RK
scheme is the fourth order explicit RK method, also known as ``RK4'' or simply
``the'' RK method. Its Butcher tableau is
\begin{equation}
  \centerline{
    \begin{tabular}{ p{1cm} | p{1cm}  p{1cm}  p{1cm}  p{1cm} }
      0 &  &  &  & \\
      $\tfrac{1}{2}$ & $\tfrac{1}{2}$ &  &  & \\
      $\tfrac{1}{2}$ & 0 & $\tfrac{1}{2}$ &  & \\
      1 & 0 & 0 & 1 & \\
      \hline
      & $\tfrac{1}{6}$ & $\tfrac{1}{3}$ & $\tfrac{1}{3}$ & $\tfrac{1}{6}$
    \end{tabular}
  }
\end{equation}
Unfortunately, this method, as any other explicit one, is not symplectic.

A particularly relevant branch of the RK family is the
Gauss-collocation scheme. Gauss-collocation methods of $s$ stages have
order $2s$, and they are both symplectic and symmetric. We chose two
of these methods for our benchmarks, specifically the second-order
``implicit midpoint rule'' (that we will call imRK2), and the
fourth-order method (imRK4).

The tableaux for imRK2 is
\begin{equation}
  \begin{tabular}{ p{1cm} | p{1cm} }
    $\frac{1}{2}$ & $\frac{1}{2}$ \\  \hline
    & 1
  \end{tabular}
\end{equation}
and leads to the non-linear equation for $\varphi(t)$
\begin{multline}
  (I + \frac{i}{2}\Delta t H\left[\overline{\varphi},t-\tfrac{1}{2}\Delta t\right]) \varphi(t) = \\
  (I - \frac{i}{2}\Delta t H\left[\overline{\varphi},t-\tfrac{1}{2}\Delta t\right]) \varphi(t-\Delta t)\,,
\end{multline}
where $\overline{\varphi} =
\tfrac{1}{2}[\varphi(t)+\varphi(t-\Delta t)]$.  Note that this
equation is similar, but not identical, to the trapezoidal or
Crank-Nicolson rule. These two methods are in fact
conjugate~\cite{Hairer1993}, and the name ``Crank-Nicolson'' is
sometimes used indistinctly for both.

The Butcher tableux for imRK4 is
\begin{equation}
  \begin{tabular}{ p{2cm} | p{2cm}  p{2cm} }
    $\tfrac{1}{2}$-$\tfrac{\sqrt{3}}{6}$ & $\tfrac{1}{4}$ & $\tfrac{1}{4}$-$\tfrac{\sqrt{3}}{6}$ \\
    $\tfrac{1}{2}$+$\tfrac{\sqrt{3}}{6}$ & $\tfrac{1}{4}$+$\tfrac{\sqrt{3}}{6}$ & $\tfrac{1}{4}$ \\
    \hline
    & $\tfrac{1}{2}$ & $\tfrac{1}{2}$
  \end{tabular}
\end{equation}

Once again, we face non-linear equations, that we implemented through
self-consistent iterative procedures similar to the ones described for
the AM formulas. Each iteration requires the solution of a linear
system. We also define ``linearized'' variants of RK as the simplified
versions in which we just perform the first step of the
self-consistent cycle with an extrapolated Hamiltonian -- a strategy
that always seems to produce the best performing algorithm.

\begin{figure}
  \centerline{
    \includegraphics[width=\columnwidth]{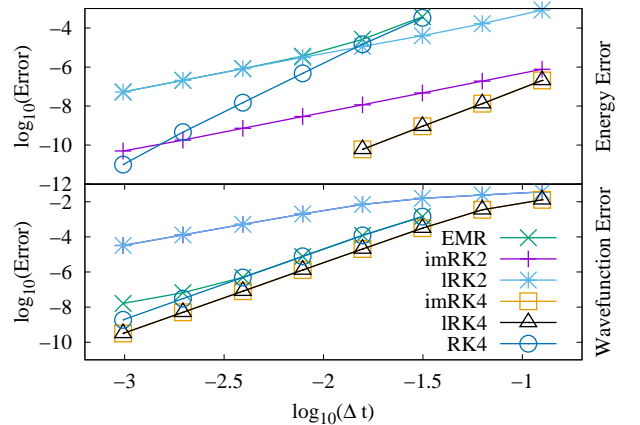}
  }
  \caption{
    \label{fig:rk1}
    Error in the total energy (top panel) and in the wave-function
    (bottom panel), as a function of the time-step, for the various RK
    methods (implicit and linearized RK2 and RK4 and explicit RK4) and
    for the EMR propagator.}
\end{figure}

We plotted the errors in the energy and wave-function as a function of the
time-step in Fig.~\ref{fig:rk1}. The points for imRK4 and lRK4 in the
energy panel that do not appear for time-steps smaller than
$\sim10^{-1.8}$ are those that reached the precision of our
machines. From the bottom panel of this figure we can see that the EMR
is significantly more precise than the second order RK methods for the
wave-function, and it can even compete with the fourth-order methods
for time-steps smaller than $10^{-2.5}$. On the other hand, as far as
the energy is concerned, we can see that EMR is outclassed by every
RK method, and especially by the implicit methods and their linearized
versions. The EMR also breaks down for the larger time-steps values,
like the RK4 method.

We also found that the linearized versions of the implicit methods
behave similarly to their full counterpart as far as the wave-function
is concerned, but that there is a significant difference in the
error of the energy (see the curves for imRK2 and lRK2).  The
explicit RK4 method performs worse both for the wave-function and
the energy when compared with the implicit methods.

\begin{figure}
  \centerline{
    \includegraphics[width=0.8\columnwidth]{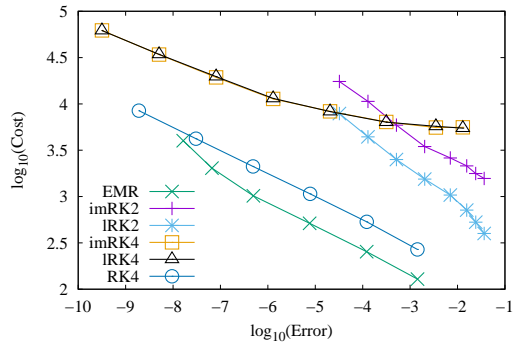}
  }
  \caption{
    \label{fig:rk2}
    Cost of the method, as a function of the error obtained (in the
    wave-function), for the various RK methods (implicit and
    linearized RK2 and RK4 and explicit RK4) and for the EMR
    propagator.  }
\end{figure}

In Fig.~\ref{fig:rk2} we show the cost as a function of the error in
the wave-function. Here the explicit methods have the advantage, with
both the EMR and RK4 performing around an order of magnitude faster than
the implicit propagators. The EMR has the best performance up to an
error of $10^{-8}$. The lRK2 performs better than the imRK2, while
imRK4 and the lRK4 have the same cost (implying that the
self-consistent cycle converged in one iteration). Among the implicit
methods, lRK2 is the best performing method up to an error of
$10^{-4}$, but for smaller values either the lRK4 or the imRK4
propagators are the best choice.

Finally, a word of caution regarding these comparisons between explicit and
implicit methods: the latter require the solution of linear systems, and their
performance will depend on the performance of the linear solvers. The
existence or not of preconditioners, for example, make these comparisons very
system and implementation dependent.

\subsubsection{Exponential RK Schemes}

Recently, we saw the appearance of the so-called ``exponential
Runge-Kutta'' (ERK)
schemes~\cite{Hochbruck2005,Maset2008,Mei2017}. The main appeal in
this family of propagators lies in its ability to tackle stiff
problems. The key idea is solving the stiff part of the equation
precisely, and approximating the remaining part by a quadrature
formula. Let us rewrite our non-linear TDKS equation as
\begin{equation}
  \varphi'(t) = -iT\varphi(t) 
-i V[\varphi(t),t]\varphi(t)\,,
\end{equation}
where $T$ is the kinetic operator (the stiff part), and the last term
is the Kohn-Sham potential acting on the orbitals. An ERK scheme for
this equation has the form
\begin{multline}
\varphi(t) = e^{-i\Delta tT}\varphi(t-\Delta t) 
\\
-i \Delta t\sum_{i=1}^s \overline{b}_i(-i\Delta t T)V[Y_i,t_i]Y_i\,,
\end{multline}
with the definition
\begin{multline}
  \label{eq:erkyi}
  Y_i = e^{-ic_i\Delta tT}\varphi(t-\Delta t) -
  \\
  i \Delta t \sum_{j=1}^s \overline{a}_{ij}(-i\Delta t T)V[Y_j,t_j]Y_j\,,
\end{multline}
Equations~\eqref{eq:erkyi} are in general a set
of $s$ non-linear equations. The constants $c_i$ and the operator
functions $\overline{a}_{ij}$ and $\overline{b}_i$ fully determine the
algorithm. These constants reduce to an \emph{underlying} RK
scheme at $T=0$, so that $\overline{a}_{ij}(0) = a_{ij}$, and
$\overline{b}_{ij}(0)=b_{ij}$. Just as with normal RK schemes, the
methods can be explicit or implicit.

For the explicit ERK schemes, we have to compute some auxiliary
functions $Y_{i}(Y_{j},t_{i})$, with $j<i$ and $i=1,...,s$, where $s$
is the number of stages of the method.  Here, the coefficients
$\overline{a}_{ij}$ and $\overline{b}_{i}$ are linear combinations of
the so-called $\phi_{k}$ functions, defined by the recurrence relation
\begin{equation}
  \phi_{k+1}(z)=\frac{\phi_{k}(z) - \phi_{k}(0)}{z},\quad\phi_{0}(z)=e^{z}
\end{equation}
leading to
\begin{equation}
  \phi_{k}(z) = \sum\limits^{\infty}_{i=0}\frac{z^{i}}{(k+i)!}.
\end{equation}
For the evaluations of the $\phi_{k}$ functions we used this Taylor
expansion. This allows us to compute both the regular exponential
function and these $\phi_{k}$, and any linear combination of them in a
simple subroutine, simplifying the implementation of the
generalization of the explicit ERK methods.

The simplest example of this family is the exponential version of the
Euler method, given by
\begin{multline}
  \varphi(t) = \varphi(t-\Delta t) + \\
  \Delta t\phi_{1}(\Delta tT)V[\varphi(t-\Delta t), t-\Delta t]\varphi(t-\Delta t)
  \,,
\end{multline}
which is an order 1 method (we call it ERK1).

We implemented a general algorithm for a broad family of ERK schemes
of any order described by Hochbruck.\cite{Hochbruck2006} We show
results for the best performing methods for orders 2, 3 and 4: method
(5.4) for order 2 (we name it ERK2), method (5.8) for order 3 (we name
it ERK3) and method (5.17) for order 4 (we name it ERK4).

Explicit exponential RK schemes cannot be symplectic (just as explicit
``normal'' RK ones), but implicit ones can.\cite{Mei2017} This is
achieved if the underlying RK method is symplectic and if the
functions $\overline{a}_{ij}$ and $\overline{b}_{i}$ obey
\begin{align}
  \overline{a}_{ij}(-i\Delta tT) =& a_{ij}e^{-i\Delta t(c_i-c_j)T}\,.
  \\
  \overline{b}_{i}(-i\Delta tT) =& b_{i}e^{-i(1-c_i)\Delta tT}\,.
\end{align}
Following this recipe, we implemented the exponential version of RK2
(labeled imERK2 in the figures), characterized by $s=1$,
$c_1=a_{11}=\tfrac{1}{2}$, and $b_1=1$, resulting in the equations
\begin{equation}
  \varphi(t) = e^{-i\Delta tT}\varphi(t-\Delta t) - i\Delta te^{-i\frac{1}{2}\Delta t T}V\left[Y,t-\frac{\Delta t}{2}\right]Y\,,
\end{equation}
with
\begin{equation}
  Y  e^{-i\frac{\Delta t}{2}T}\varphi(t-\Delta t) - i \frac{\Delta t}{2}V\left[Y,t-\frac{\Delta t}{2}\right]Y\,.
\end{equation}

\begin{figure}
  \centerline{
    \includegraphics[width=\columnwidth]{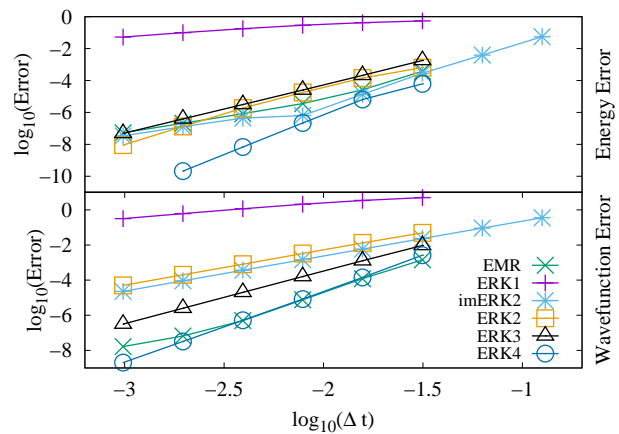}
  }
  \caption{
    \label{fig:exprk1}
    Error in the total energy (top panel) and in the wave-function
    (bottom panel), as a function of the time-step, for the various
    exponential RK methods (exponential Euler method, implicit RK2 and
    explicit RK2, RK3 and RK4) and for the EMR propagator.}
\end{figure}

\begin{figure}
  \centerline{
    \includegraphics[width=0.8\columnwidth]{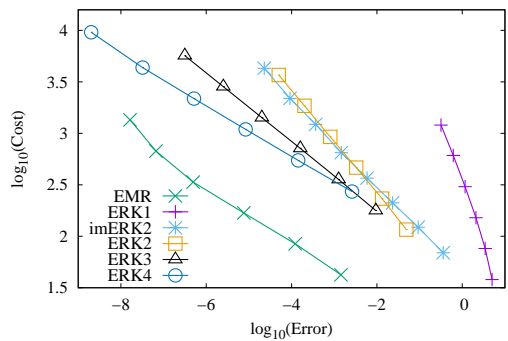}
  }
  \caption{
    \label{fig:exprk2}
    Cost of the method, as a function of the error obtained (in the
    wave-function), for the various exponential RK methods
    (exponential Euler method, implicit RK2 and explicit RK2, RK3 and
    RK4) and for the EMR propagator.}
\end{figure}

Figs.~\ref{fig:exprk1} and \ref{fig:exprk2} display the numerical
results obtained for this implicit method, and for the four explicit
methods mentioned above, compared to the EMR scheme. As we can see,
all the exponential methods keep their order for the values of the
time-step where they do not break down. Furthermore, the implicit
version of the exponential RK2 propagator has a wider range of
stability than the explicit versions. From the top panel we can see
that imERK2 has slightly smaller errors for the energy than EMR2
up to $\Delta t\sim 10^{-2.7}$, and it is always better than the EMR
as far as the error in the energy is concerned. The ERK4 beats every
other propagator in the top panel.  On the other hand, for the
wave-function the EMR is on par with ERK4, being slightly more
precise for the largest values of the time-step, and only being
overtaken by ERK4 when $\Delta t<10^{2.5}$. If we compare the explicit
and implicit ERK2, we can see that imERK2 has a smaller error in the
wave-function than ERK2, but in the energy comparison ERK2 is better
for time-step values below $10^{-2.5}$. As we have mentioned before,
this figure clearly shows that the methods behave as expected from the
theoretical formulas, maintaining their order during the whole range
of time-steps studied.

From Fig.~\ref{fig:exprk2} we can see that these methods are
computationally expensive, with none of the exponential RK methods
coming close to the EMR cost. Among the ERK family, ERK4 has the best
performance for values of the error in the wave-function below
$10^{-2.2}$, making it the best overall ERK method from the ones we
tested.

\section{Conclusions}
\label{sec:conclusions}

We implemented and analyzed four families of numerical integrators for
the Kohn-Sham equations in our code octopus, specifically
commutator-free Magnus expansions, multistep methods, Runge-Kutta
propagators and exponential Runge-Kutta integrators. These were
compared to the previously studied exponential mid-point rule,
enforced time-reversal symmetry and approximately enforced
time-reversal symmetry propagators. For each method we evaluated the
error in the wave-function and the energy as a function of the
time-step, together with the cost in computational time as a function
of the error.

Among the new families of propagators studied in this paper, the
fourth-order commutator-free Magnus expansion beats every other
propagator in terms of cost/accuracy, making it the recommended method
for TDDFT. The multistep integrators main advantage is 
that the computational cost remains constant independently of the
number of previous steps considered. The exponential Runge-Kutta
propagators do not show any clear advantage over the regular
Runge-Kutta methods, with the explicit Runge-Kutta
method of fourth-order being usually the best choice. The exception are
stiff problems or in situations where a high degree of conservation of
some quantity is required. In such cases, the symplecticity of the
implicit versions of Runge-Kutta comes into play.

We have shown how the TDKS equations, in the adiabatic approximation, form a
Hamiltonian and therefore a symplectic ODE system. Therefore, for long time
propagations one should benefit from the use of structure preserving
algorithms. This fact discourages the use of multistep schemes, for example,
and favors implicit schemes that are unfortunately less
cost-effective.

The numerical integration of first-order ordinary differential
equations is a very active field of research, with new schemes being
proposed and old ones refined regularly, and many other methods 
still untested. We can therefore still expect new developments in the
numerical propagation of the time-dependent Kohn-Sham equations,
opening the way for the study of larger systems for longer periods of
time.


%
%

%

\begin{acknowledgement}
We acknowledge support from Ministerio de Econom{\'{\i}}a y
Competitividad (MINECO) grants FIS2013-46159-C3-2-P and
FIS2014-61301-EXP, from the European Research Council
(ERC-2015-AdG-694097), from Grupos Consolidados (IT578-13), from the
European Union Horizon 2020 program under Grant Agreement 676580
(NOMAD), from the Salvador de Madariaga mobility grant
PRX16/00436, and from the DFG Project B09 of TRR 227.
\end{acknowledgement}

\bibliography{article}

\end{document}